# NEW HYPERGEOMETRIC SERIES SOLUTIONS TO THE GENERAL HEUN EQUATION

R.S. SOKHOYAN[*], D.Yu. MELIKDZANIAN and A.M. ISHKHANYAN[**]

**We introduce new hypergeometric series expansions of the solutions to the general Heun equation. The form of the Gauss hypergeometric functions used as expansion function differs from that used before. We derive three such expansions and further generate, by termination of the series, closed-form solutions for several sets of involved parameters.**

**INTRODUCTION**

The Heun functions [1] satisfying the following second-order linear differential equation with four regular singular points located at $z = 0, 1, a$ and $\infty$

$$u'' + \left(\frac{\gamma}{z} + \frac{\delta}{z-1} + \frac{\varepsilon}{z-a}\right)u' + \frac{\alpha\beta\, z - q}{z(z-1)(z-a)} u = 0 \qquad (1)$$

(where the parameters satisfy the Fuchsian condition $1 + \alpha + \beta = \gamma + \delta + \varepsilon$ to ensure the regularity of the singular point at infinity) and the ones obeying the four confluent Heun equations generated by coalescence of singularities of this equation [2,3] are believed to become a part of the next generation of standard special functions of mathematical physics in the near future. Equation (1) is the most general Fuchsian second-order linear differential equation with four singular points in the sense that any such equation with exactly four singular points can be transformed to the Heun equation by appropriate changes of the independent and dependent variables.

Since the general Heun equation (1) presents a natural generalization of the Gauss hypergeometric equation [4] by having one more regular singular point, the equations of the Heun class are widely faced in current frontline physics and mathematics research; the occurrences of such equations are too numerous to be mentioned in detail (for a survey, see [2]). However, the Heun equations are still much less studied than those of the hypergeometric class; quoting [2] (page 4): "the difficulties which occur when attacking these equations are greater by at least an order of magnitude than those we encounter when solving the hypergeometric equation". For this reason, the solutions in terms of mathematical functions that satisfy less complicated differential equations are of considerable interest.

However, exact solutions of the Heun equations in terms of simpler mathematical functions (in particular, hypergeometric ones) are very rare. A possible way to obtain such solutions is to transform the independent and dependent variables in order to reduce the equations to simpler ones [5, 6]. The significant potential of this approach has been recently demonstrated by Maier, who studied Heun-to-hypergeometric reductions via polynomial (quadratic, cubic, etc.) and several rational transformations of the variables [5].

[*]Email: sruzan@gmail.com
[**]Email: aishkhanyan@gmail.com

However, the most *systematic* method for derivation of exact solutions to the Heun equations is the termination of a series expansion (in fact, this gives the largest set of known closed-form solutions). Such expansions were first discussed by Svartholm and Erdélyi [7]. Later on, this technique has been applied to the Heun equation (1) and its confluent forms by numerous authors. Expansions in terms of Gauss hypergeometric and confluent hypergeometric functions have been constructed and applied to a number of problems [2].

In the present paper, we introduce new expansions of the solutions to the general Heun equation (1) in terms of Gauss hypergeometric functions. We show that the expansions generate closed-form explicit solutions for several infinite sets of specific values of involved parameters. Since the explicit solutions are generated through a regular, systematic algorithm, they can be easily incorporated into a computer symbolic algebra system.

**EXPANSIONS**

We introduce a solution of Eq. (1) of the form

$$u = \sum_n a_n u_n = \sum_n a_n \cdot {}_2F_1(\alpha, \beta; \gamma_0 - n; z) \tag{2}$$

with involved Gauss hypergeometric functions obeying the equation

$$u_n'' + \left( \frac{\gamma_0 - n}{z} + \frac{\delta + \varepsilon + \gamma - \gamma_0 + n}{z - 1} \right) u_n' + \frac{\alpha \beta}{z(z-1)} u_n = 0. \tag{3}$$

Substitution of Eq. (2) into Eq. (1) then gives

$$\sum_n a_n \left[ \left( \frac{\gamma - \gamma_0 + n}{z} - \frac{\varepsilon + \gamma - \gamma_0 + n}{z - 1} + \frac{\varepsilon}{z - a} \right) u_n' + \frac{\alpha \beta a - q}{z(z-1)(z-a)} u_n \right] = 0. \tag{4}$$

or

$$\sum_n a_n \{ [(a-1)(\varepsilon + \gamma - \gamma_0 + n)z - a(\gamma - \gamma_0 + n)(z-1)] u_n' + (\alpha \beta a - q) u_n \} = 0. \tag{5}$$

Now, using the following relations between hypergeometric functions

$$z \cdot \frac{d\,{}_2F_1(\alpha, \beta; \gamma; z)}{dz} = (\gamma - 1)[{}_2F_1(\alpha, \beta; \gamma - 1; z) - {}_2F_1(\alpha, \beta; \gamma; z)], \tag{6}$$

$$(z-1) \cdot \frac{d\,{}_2F_1(\alpha, \beta; \gamma; z)}{dz} = -(\alpha + \beta - \gamma) \cdot {}_2F_1(\alpha, \beta; \gamma; z) + \left( \alpha + \beta - \gamma - \frac{\alpha \beta}{\gamma} \right) \cdot {}_2F_1(\alpha, \beta; \gamma + 1; z), \tag{7}$$

so that:

$$z \cdot u_n' = \gamma_{n+1}[u_{n+1} - u_n], \tag{8}$$

$$(z-1) \cdot u_n' = -\delta_{n-1} u_n + \left( \delta_{n-1} - \frac{\alpha \beta}{\gamma_n} \right) u_{n-1} \tag{9}$$

with $\gamma_n = \gamma_0 - n$ and $\delta_n = \delta + \varepsilon + \gamma - \gamma_0 + n$, this equation is straightforwardly rewritten as

$$\sum_n a_n \big[ (a-1)(\varepsilon+\gamma-\gamma_0+n)(\gamma_n-1)[u_{n+1}-u_n]$$
$$+ a(\gamma-\gamma_0+n)\left( (\delta_n-1)u_n - \left(\delta_{n-1}-\frac{\alpha\beta}{\gamma_n}\right)u_{n-1}\right) + (\alpha\beta a-q)u_n \big] = 0, \quad (10)$$

from which we get a three-term recurrence relation for the coefficients of the expansion (2):

$$R_n a_n + Q_n a_{n-1} + P_n a_{n-2} = 0 \quad (11)$$

with

$$R_n = -a(\gamma-\gamma_0+n)\left(\delta+\varepsilon+\gamma-\gamma_0+n-1-\frac{\alpha\beta}{\gamma_0-n}\right), \quad (12)$$

$$Q_n = -(a-1)(\varepsilon+\gamma-\gamma_0+n-1)(\gamma_0-n) + a(\gamma-\gamma_0+n-1)(\delta+\varepsilon+\gamma-\gamma_0+n-2) + (\alpha\beta a-q), \quad (13)$$

$$P_n = (a-1)(\varepsilon+\gamma-\gamma_0+n-2)(\gamma_0-n+1). \quad (14)$$

Finally, in order to terminate the expansion from the left-hand side, we put $a_{-2}=a_{-1}=0$ and demand $a_0$ be an arbitrary constant. Hence, should be $R_0=0$ and we get from Eq. (12) that necessarily

$$\gamma_0 = \gamma \text{ or } \alpha \text{ or } \beta. \quad (15)$$

Thus, we have constructed three expansions of the solutions to the Heun equation in terms of hypergeometric functions having the form $_2F_1(\alpha,\beta;\gamma_0-n;z)$ with $\varepsilon+\gamma-\beta=-N$. For any set of parameters of the Heun equation, provided that $\gamma,\alpha,\beta$ all simultaneously are not integers, at least one of these expansions can be applied. The functions applied above differ from that used by Svartholm, Erdelyi and Schmidt [7] in the earlier discussions of hypergeometric-function expansions; they used functions of the form $_2F_1(\lambda+n,\mu-n;\gamma;z)$ (see a summary on this topic by Arscott in [2]). The expansions also differ from the Jacobi-polynomial-expansion constructed by Kalnins and Miller whose functions can be rewritten in terms of functions of the form $_2F_1(\lambda+n,\mu-n;\nu+2n;z)$ [8].

Here we are particularly concerned with the cases when the expansions terminate thus resulting in closed-form solutions. Evidently, this occurs when any two successive coefficients of expansion (2) are equal to zero. Let $a_n$ be the last non-zero coefficient: $a_n \neq 0$, $a_{n+1}=0$ and $a_{n+2}=0$, for some $n=N\geq 0$. This results in two conditions imposed on the parameters of the Heun equation. First, from Eq. (11) written for $n=N+2$ we conclude that $P_{N+2}=0$ and hence, should be

$$\varepsilon, \ \varepsilon+\gamma-\alpha \text{ or } \varepsilon+\gamma-\beta = -N \quad (16)$$

for the first, second and third expansions, respectively ($N=0,1,2,\ldots$). The second condition is convenient to write in terms of an infinite fraction:

$$Q_1 - \cfrac{R_1 P_2}{Q_2 - \cfrac{R_2 P_3}{Q_3 - \cfrac{R_3 P_4}{Q_4 - \ldots}}} = 0. \qquad (17)$$

Substituting here Eq. (16) terminates the fraction and thus leads to a polynomial equation for the accessory parameter $q$ of the order of $N+1$. Hence, in general, for any given $N$ there exist $N+1$ cases for which expansion (2) terminates. The solution to the Heun equation is then a linear combination of $N+1$ hypergeometric functions.

In the case $\gamma_0 = \gamma$ the involved hypergeometric functions have the form $u_n = {_2F_1}(\alpha, \beta; \gamma - n; z)$. Solutions consisting of finite sums of functions of this form were first derived by Craster and co-workers [9]. Thus, the infinite expansion (2) is a direct generalization of the idea by Craster. The termination condition for this case is $\varepsilon = -N, N = 0, 1, 2, \ldots$. Using Eqs. (11)-(17) it is not difficult to write down the finite-sum solutions and corresponding equations for the accessory parameter in explicit form. Here are the first three of them (one-, two- and three-term solutions) (compare with [9]):

$$\varepsilon = 0 \text{ (trivial case)}, \qquad (18)$$

$$q - a\alpha\beta = 0, \qquad (19)$$

$$u = {_2F_1}(\alpha, \beta; \gamma; z), \qquad (20)$$

$$\varepsilon = -1, \qquad (21)$$

$$(q - a\alpha\beta)^2 + [(1-a)(\gamma-1) + a(1-\delta)](q - a\alpha\beta) - a(1-a)\alpha\beta = 0, \qquad (22)$$

$$u = {_2F_1}(\alpha, \beta; \gamma; z) + \frac{(\gamma-1)(q - a\alpha\beta + (1-a)(\gamma-1))}{a(\alpha\beta - (\gamma-1)(\delta-1))} \cdot {_2F_1}(\alpha, \beta; \gamma-1; z), \qquad (23)$$

$$\varepsilon = -2, \qquad (24)$$

$$x^3 + [3\gamma + a(-3\gamma - 3\delta + 8) - 4]x^2$$

$$+ 2[2 - 3\gamma + \gamma^2 - 2a(\alpha\beta + (\gamma-1)(\gamma+\delta-3)) + a^2(2\alpha\beta + (\gamma+\delta-3)(\gamma+\delta-2))]x$$

$$-4a(a-1)\alpha\beta[-\gamma + a(\gamma+\delta-2) + 1] = 0, \qquad (25)$$

$$u = {_2F_1}(\alpha, \beta; \gamma; z) + a_1 \cdot {_2F_1}(\alpha, \beta; \gamma-1; z) + a_2 \cdot {_2F_1}(\alpha, \beta; \gamma-2; z), \qquad (26)$$

where $x = q - a\alpha\beta$, $a_1 = -Q_1/R_1$, $a_2 = -(Q_2 a_1 + P_2)/R_2$.

The second independent solution complementary to these solutions is given by means of the expansion

$$u = \sum_{n=0}^{\infty} a_n \cdot {_2F_1}(\alpha, \beta; \gamma_0 - n; 1-z), \qquad (27)$$

where for $\gamma_0$ stands

$$\gamma_0 = \delta, \qquad (28)$$

and the coefficients $a_n$ are defined by the same recurrence relations (11)-(14) with parameters redefined as $\gamma \leftrightarrow \delta$, $\gamma_0 \to \delta$, $a \to 1-a$, $q \to -q+\alpha\beta$. To see this, first note that the independent variable transformation $z = 1-x$ transforms the initial Heun equation (1) to the form

$$u'' + \left(\frac{\delta}{x} + \frac{\gamma}{x-1} + \frac{\varepsilon}{x-(1-a)}\right)u' + \frac{\alpha\beta x - (-q+\alpha\beta)}{x(x-1)(x-(1-a))}u = 0, \qquad (29)$$

so that above expansion (2) can be applied to this equation as well. And expansion (27) with (28) is just the one. Furthermore, importantly, it can be checked that expansion (27) terminates for exactly the same values of the accessory parameter $q$ as the previous expansion (2) does. Hence, the expansions (2) and (27) produce two linearly independent finite-sum solutions for the same set of the parameters of the initial Heun equation (1).

As regards to the finite-sum solutions produced by the choices $\gamma_0 = \alpha$ and $\beta$, these two are identical solutions because of $\alpha \leftrightarrow \beta$ interchange symmetry of equation (1). The involved hypergeometric functions here have the form $u_n = {}_2F_1(\alpha,\beta;\alpha-n;z)$ [or ${}_2F_1(\alpha,\beta;\beta-n;z)$] so that are reduced to simpler functions. The functions can be rewritten in terms of Jacobi polynomials with specified parameters. Applying the formula [4]

$$_2F_1(\alpha,\beta;\alpha-n;z) = (1-z)^{-\beta-n} \cdot {}_2F_1(-n,-\beta+\alpha-n;\alpha-n;z), \qquad (30)$$

it can be shown that the finite-sum can eventually be written as a product of $(1-z)^{1-\delta}$ and a polynomial in $z$. Here are the first two of the solutions:

$$\varepsilon + \gamma - \alpha = 0 \quad (\Leftrightarrow \quad \beta = \delta - 1), \qquad (31)$$

$$q - a\gamma(\delta-1) = 0, \quad u = (1-z)^{1-\delta}, \qquad (32)$$

$$\varepsilon + \gamma - \alpha = -1 \quad (\Leftrightarrow \quad \beta = \delta - 2), \qquad (33)$$

$$q^2 + [\gamma - a(\delta-2+\gamma(2\delta-3)) + \varepsilon]q - a\gamma(\delta-2)[(1+\gamma+\varepsilon) - a(1+\gamma)(\delta-1)] = 0, \qquad (34)$$

$$u = (1-z)^{1-\delta}\left(1 - \frac{2+\gamma-\delta+\varepsilon}{\gamma+\varepsilon}z + \frac{q-a(\alpha\beta+\varepsilon-\delta\varepsilon)}{(1-a)(\gamma+\varepsilon)}(1-z)\right). \qquad (35)$$

Note, finally, that the transformation $u = (1-z)^{1-\delta}v(z)$ results in a Heun equation for $v(z)$ with parameters $\gamma_1 = \gamma$, $\delta_1 = 2-\delta$, $\varepsilon_1 = \varepsilon$, $\alpha_1\beta_1 = \alpha\beta - (\delta-1)(\gamma+\varepsilon)$ and $q_1 = q - a\gamma(\delta-1)$. Hence, the cases $\varepsilon+\gamma-\alpha = -N$ (and $\varepsilon+\gamma-\beta = -N$) can be viewed as some polynomial-solution cases of the equation for $v$, the case (31)-(32) standing for a trivial solution corresponding to missed $v$-term case when $\alpha_1\beta_1 = 0$ and $q_1 = 0$.

**DISCUSSION**

Thus, we have discussed a new type of the Gauss hypergeometric function expansions of the solutions of the general Heun equation. We have shown that if $\gamma, \alpha, \beta$ are not integers, the solutions of the Heun equation admit expansions of the form

$$u = \sum_{n=0}^{\infty} a_n \cdot {}_2F_1(\alpha, \beta; \gamma_0 - n; z) \text{ with } \gamma_0 = \gamma, \alpha, \beta. \tag{36}$$

The expansions generate closed-form solutions in three cases: $\varepsilon$, $\varepsilon + \gamma - \alpha$, $\varepsilon + \gamma - \beta = -N$, $N = 0, 1, 2, 3, \ldots$. In each case the general Heun equation admits closed form solutions at, in general, $N+1$ choices of the accessory parameter $q$ defined by a polynomial equation of the order of $N+1$. The most non-trivial case is the case of negative integer $\varepsilon$, $\varepsilon = -N$, when the solutions involve $N+1$ hypergeometric functions (in general) irreducible to simpler functions. Further, since for any *positive* integer $\varepsilon = +N \geq 2$ the transformation $u = (z-a)^{1-\varepsilon} v(z)$ leads to a Heun equation with a zero or negative integer exponent $\varepsilon_1 = 2 - \varepsilon \leq 0$:

$$v'' + \left(\frac{\gamma}{z} + \frac{\delta}{z-1} + \frac{2-\varepsilon}{z-a}\right)v' + \frac{(\alpha\beta - (\varepsilon-1)(\gamma+\delta))z - (q - \gamma(\varepsilon-1))}{z(z-1)(z-a)} v = 0, \tag{37}$$

by applying above expansions to this equation we are able to construct expansions involving hypergeometric functions of the form ${}_2F_1(\alpha, \beta; \gamma_0 - n; z)$ also for positive integer $\varepsilon = +N \geq 2$. (In this case the equation for $q$ is of the order of $N-1$ and the expansion consists of $N-1$ hypergeometric functions.) Thus, the only uncovered case among integer $\varepsilon s$ remains $\varepsilon = 1$.

The derived solutions together with another major set that is generated by terminating the Beta-function expansions presented in our earlier paper [10] amount the largest set of closed form solutions known at present time for the Heun equation. The explicit solutions are generated in a simple way, using a regular, systematic algorithm that can be easily incorporated into a computer symbolic algebra system such as Mathematica and Maple.

Finally, it should be mentioned that, evidently, analogous expansions can be constructed for other equations including those of the confluent Heun class. For example, one could try to expand the solutions of the singly confluent Heun equation using the Kummer confluent hypergeometric functions of the form ${}_1F_1(\alpha, \gamma_0 - n; z)$. Further possibilities may also open when one uses as expansion functions *combinations* of these functions, as well as preliminarily transforms the initial equation via change of the independent and dependent variables (examples of such developments have been constructed in our recent paper [11]). We will consider these possibilities in future publications.

This work was supported by the International Science and Technology Center (Grant A-1241) and Armenian National Science and Education Fund (Grants No. PS-10-2005 and No. PS-11-2005).